# Microvalve-Based Tunability of Electrically Driven Ion Transport Through a Microfluidic System with Ion-Exchange Membrane

Barak Sabbagh[1], Sinwook Park[2], Gilad Yossifon[1,2*]

[1]Faculty of Mechanical Engineering, Technion–Israel Institute of Technology, Israel

[2]School of Mechanical Engineering, Tel-Aviv University, Israel

**ABSTRACT**

Microfluidic channels with embedded ion permselective medium under the application of electric current are commonly used for electrokinetic processes as on-chip ion concentration polarization (ICP) and bioparticle preconcentration to enhance biosensing. Herein, we demonstrate the ability to dynamically control the electrically driven ion transport by integrating individually addressable microvalves. The microvalves are located along a main microchannel that is uniformly coated with a thin layer of an ion-exchange membrane (IEM). The interplay of ionic transport between the solution within the microchannel and the thin IEM, under an applied electric current, can be locally tuned by the deformation of the microvalve. This tunability provides a robust and simple means of implementing new functionalities into lab-on-a-chip devices, e.g., dynamic control over multiple ICP layers and their associated preconcentrated molecule plugs, multiplex sensing, suppression of biofouling as well as plug dispersion, while maintaining the well-known application of microvalves as steric filtration.

**KEYWORDS:** nanofluidics, ion permselectivity, ion transport, ion concentration polarization, microvalve

## INTRODUCTION

Passage of an electric current through an ionic permselective medium (e.g., nanochannel or ion-exchange membrane) results in an electrokinetic phenomenon termed ion concentration-polarization (ICP). Under application of a direct current (dc) electric field, the symmetry-broken transport of ions through the permselective medium triggers ionic depletion and enrichment diffusion layers at the two opposite interfaces of the permselective medium[1,2]. This then leads to a non-linear current-voltage response consisting of three regimes. The first is of a linear Ohmic-like response (i.e., under-limiting regime) that later transitions to a plateau-like response (limiting regime) at the limiting current, and upon further increase of the voltage it shifts to an over-limiting current that continuously increases (over-limiting regime)[3,4]. The interplay between the various microchannel-related resistances and the ionic permselective medium related resistance determines the limiting current ($I_{Tot}^{Lim}$)[5]. The transition to the limiting regime occurs due to the increased electrical resistance associated with the vanishing of the electrolyte ionic concentration at the depleted membrane-microchannel interface.

This ion depletion results in a strong electric field gradient, which traps and preconcentrates charged bioparticles into a plug through a mechanism known as field gradient focusing [6,7]. The trapping occurs due to a force balance between the counter-acting advection and electro-migration at the edge of the depletion layer. The concentration of the target bioparticle (e.g., DNA[8,9], protein[10,11] and bacteria[12,13]) at the plug can reach several orders of magnitude of the initial concentration, which significantly enhances its detection. Among other electrokinetic-based bioparticle preconcentration techniques, e.g. dielectrophoresis[14,] and isotachophoresis[15] molecular trapping, ICP-driven preconcentration is regarded as one of the most efficient and common tools for enhancing the detection of charged bioparticles in microscale bioanalysis[16]. ICP-driven preconcentration is commonly realized within microfluidic channels with relatively high hydraulic permeability, either by employing the ionic permselective medium as a bridge between two microchannels[10,17,18] or as a patterned thin surface coating embedded at the bottom of the main microchannel[13,19–22]. These microfluidic system designs maintain minimal hydrodynamic interference within the microfluidic channels and support sufficiently high flow rate and flux of target bioparticles towards the plug. However, for robust preconcentration, aside from the requirement for high throughput to achieve rapid bioparticle accumulation, precise overlap of the plug with the sensing region (e.g., immobilized molecular probes as antibodies[10,23], or electrodes for electrochemical sensing[9]) is essential. One way of achieving such an overlap is via extensive precalibration involving an elaborate process of trial and error to define the optimal operation conditions (e.g., applied voltage, flow rate) as a function of the system parameters (e.g., ionic strength, geometry, target molecules). Microfluidic channel geometry variations can assist in localizing the plug, although some precalibration is still required[24,25].

A more direct and precise approach involves the active control of plug location via powered electrodes. The electrodes are embedded within the microchannel for localized stirring of the fluid, driven by either alternating current electro-osmosis (ACEO)[26] or electrothermal[27] flow. These methods require electrode fabrication and are mostly limited to a relatively low ionic strength electrolyte (< 2mM). Recently, we developed a series of tunable nano-channels by using microvalves made of a soft elastomer (PDMS)[28]. Each microvalve enables tuning of the cross-sectional area of the main microchannel from micro- to nano- meter scale, thereby switching the microchannel into an ionic perm-selective nanochannel[28–30]. It was shown that a series of such tunable nanochannels can be used to generate multiple plugs in series, wherein the location of the multiple plugs can be dynamically controlled based on which microvalves were operated. However, the fact that the solution must flow through nanochannels with high hydrodynamic resistances results in a low throughput and correspondingly prolonged process, with a low preconcentration factor.

The present work combined the advantages of ICP using a thin cation exchange membrane (CEM) coating (acting as the ionic permselective medium), with the advantages of a tunable microchannel geometry achieved by utilizing multiple elastomeric microvalves. The use of microvalves eliminates the need to pattern the CEM during the chip fabrication process. Instead, a uniform CEM coating was deposited along the entire bottom surface of the microchannel. Despite the uniform CEM coating, a highly controlled ICP-driven preconcentration plug was dynamically formed around each individual microvalve. When a fixed voltage drop was applied from both ends of the microchannel with a sufficiently deformed microvalve, the interplay of ionic transport between the solution and the CEM led to generation of ICP and preconcentration plug. Although the deformed microvalve did not fully block the microchannel, it could deflect sufficient ionic current through the CEM to trigger the ICP, while the remaining gaps of several microns allowed for moderate hydraulic permeability. This behavior was examined and optimized under various operation conditions (e.g., microvalve deformations, ionic strengths, and applied voltages) and verified with a numerical model. Integration of an array of individually addressable microvalves enabled control of the location of both the ICP region and the preconcentration plugs via interaction between multiple ICP regions. In contrast to other studies that were based on fixed system geometry and properties, the unique device presented here opens new opportunities, including real-time tuning of the overall performance, in particular the system's ionic permselectivity, spatio-temporal control of ICP and multiple preconcentration plugs.

## RESULTS AND DISCUSSION

### *Working principle*

The system response roughly divided into three modes, defined by the deformation level of the microvalve and the corresponding ICP response (i.e., Ohmic, ICP with/without an effective formation of a plug) (Fig.1). When a relatively low pressure is applied within the control channel ($P_I$), the microvalve is considered to be deactivated (i.e., not deformed) and the main microchannel, tens of microns in height, remains open (Mode [I]). Despite the deactivated microvalve, the uniform CEM coating on the microchannel bottom surface causes the electrically driven ion transport that supports the electrical current ($I_{Tot}$) to divide between the CEM ($I_{CEM}$) and the bulk solution ($I_{Sol}$). The ratio between the two currents is determined by the electrical resistance of each component. A simplified equivalent circuit of the system's response for the initial times after application of an external electric potential/current (previously to the possible emergence of diffusion layers) was used to describe this behavior (Fig.1). The circuit consists of fixed anodic and cathodic microchannel resistors ($R_{Anod}$, $R_{Cath}$) external to the microvalve section. These are connected in series to the resistances within the microvalve section comprised of a geometry-dependent variable resistor of the solution, $R_{Sol}$, and a fixed CEM resistor, $R_{CEM}$, which are connected in parallel. Hence, in the case of a deactivated microvalve, the ratio of $I_{CEM}/I_{Sol}$ is uniform along the entire channel due to the uniform cross-section geometry. If the electrical current passing through the edge of the CEM (closest to the anode electrode) is insufficient to generate ICP, there will be no generation ICP along the entire microchannel (i.e., $I_{Tot} < I_{Tot}^{Lim}$). Unlike the electrical current, the solution fluid flow between the two inlets is through the main channel, due to the negligible hydraulic permeability of the CEM. A further increase of P deforms the microvalve, wherein at a certain threshold ($P_{II}$), the deformation is sufficient to divert enough electrical current through the CEM and formation of a stable ICP that exceeds $I_{Tot}^{Lim}$ (Mode [II], $I_{Tot} > I_{Tot}^{Lim}$). At this state, the valve deformation partially blocks the cross-sectional area of the microchannel while leaving gaps of several microns. This reduced cross-section modifies the local interplay between $I_{Sol}$ and $I_{CEM}$ by increasing $R_{Sol}$ (while $R_{CEM}$ is unaffected), which, in turn, leads to increased $I_{CEM}/I_{Tot}$. Yet, these remaining gaps enable sufficient hydraulic permeability for moderate fluid flow. An even further increase of P ($P_{III}$) deforms the microvalve such that these gaps are further reduced to sub-micron sizes which substantially increase the hydrodynamic resistance (Mode [III]). For the purpose of developing an effective preconcentration plug with a high concentration factor, conditions of sufficient ion transport through an ion permselective medium together with comparatively high hydraulic permeability are essential. Mode II meets these requirements and thereby efficiently generates a plug with a high accumulation factor.

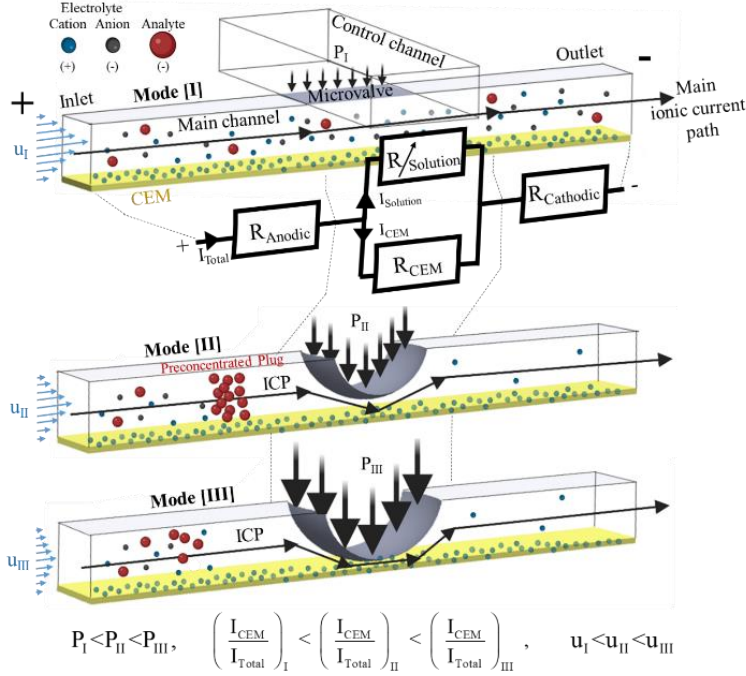

**Figure 1: Schemes of the three modes of the system electrically driven ion transport response for varying control channel pressure (P).** Mode [I]-Ohmic, [II]- ion concentration-polarization (ICP) with a preconcentrated analyte plug, and [III]-ICP with a negligible plug. The microvalve is located at the overlapping region (marked in grey) between the main microchannel and the control channel, while the cation exchange membrane (CEM) (marked in yellow) is on the bottom surface of the main channel. The electrolyte cations and anions, and the analyte are represented by blue, black, and red spheres, respectively. The black and blue arrows qualitatively represent the main electrical current path (black) and flow velocities (blue). The simplified equivalent circuit that describes the system's response consists of two resistances external to the microvalve region ($R_{Anod}$, $R_{Cath}$), and two within the microvalve region ($R_{Sol}$, $R_{CEM}$). $R_{Sol}$ is a variable resistor (marked with an arrow), as opposed to all other resistors which are of a fixed Ohmic value.

An experimental demonstration of on-demand local application of ICP and the development of the corresponding preconcentration plug by microvalve activation is shown in Figure 2. At times 0-10s, despite the applied electric potential ($\phi$=60V), when the microvalve was not active (P=0psi, Mode [I]), the system did not exhibit any visual indication of development of ICP (i.e., uniform fluorescence intensity). Only upon deformation of the microvalve (t>10s, P=16psi, Mode [II]), did fluorescent molecules accumulate into a highly preconcentrated plug at the edge of the depletion layer. The plug propagated towards the anodic side as the depletion layer grew in size. Deactivation of the microvalve (P=0 psi, Mode [I]) resulted in immediate flushing downstream of the plug and with it, any trace of the ICP, despite the continuous application of the electric field. The experimental results were confirmed with numerical simulations (Fig. 2, right column and Supp. M1).

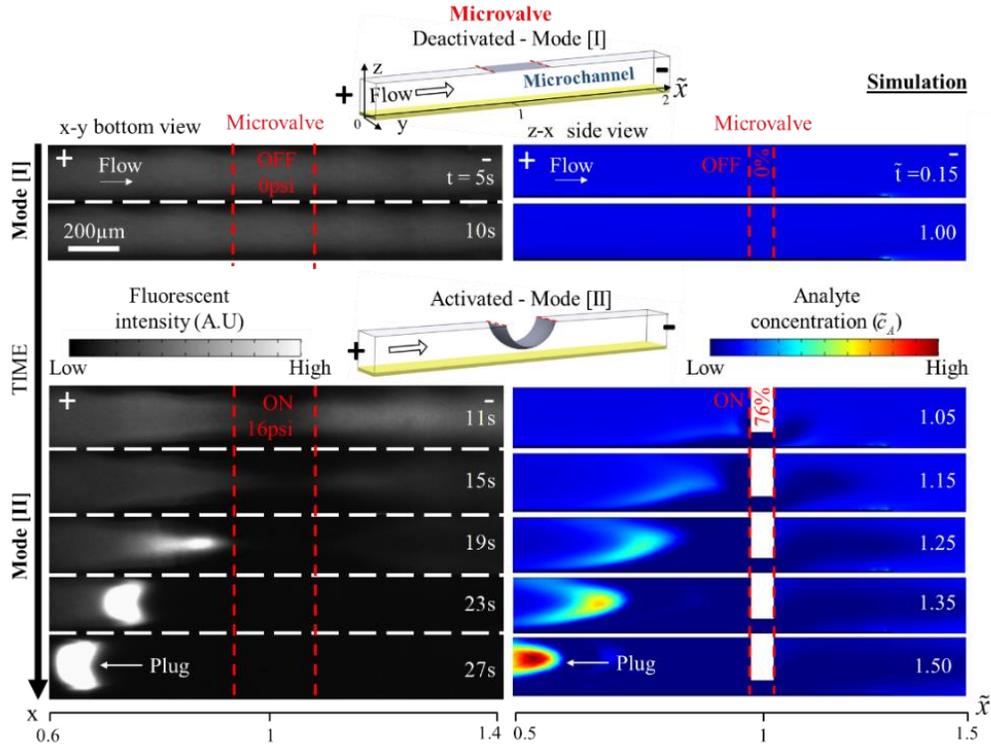

**Figure 2: On-demand generation of ion concentration-polarization (ICP) and a preconcentrated analyte plug by a single microvalve activation**. Experimental and numerical results (right and left columns, respectively) upon activation of a microvalve (Mode [I] → Mode [II]) under a continuously applied voltage drop within the main microchannel. The experimentally measured fluorescence intensity (arbitrary units, A.U) and numerically computed normalized analyte concentration ($\tilde{c}_A$) over time are shown. The microvalve region is marked by two dashed red lines. Experiments: bottom-view of channel, $P_{Mode[I]}$=0psi, $P_{Mode[II]}=16 psi$, $\phi$=60V, u=270μm/sec (velocity obtained with an open channel). Simulations: side-view of channel, 0% and 76% closures, $\tilde{N}=8$, $\tilde{\phi}=85$ and ũ=8.

*ICP dependency on the microvalve deformation – experimental investigation*

The system response was experimentally studied via visualization and electrical characterization of various microvalve deformations by gradually increasing P [Fig.3][Supp.M2]. Confocal imaging-based three-dimensional reconstruction of the cross-section shape of the main microchannel beneath the activated microvalve revealed non-uniform closure, with maximum deformation at the center of the microvalve and gaps at the sides of the cross-sectional area through which the solution can pass [Fig.3A]. This deformation evolves from the initial curved cross-sectional shape of the main channel [Fig.S2-3]. To simplify the analysis, instead of considering the precise gap geometry, the approximated closure percentage of the main microchannel underneath the microvalve was considered. As discussed already, there was no visual indication (i.e., no changes in the fluorescence intensity) of ICP when the microvalve was deactivated (P=0psi) [Fig.3B]. In agreement, the chronoamperometric response (i.e., electric current resulting from a step-wise application of a constant voltage drop) exhibited a steady current over time ($I_{Tot}$~2.75µA) that was proportional to the electrolyte Ohmic resistance as occurs in non-selective systems, justifying the negation of an induced ICP- related diffusion layers contribution to the resistance [Fig.3C-E]. Increasing P beyond 18psi switched the system behavior to a non-linear I-V response (Mode [II]) typical to ion permselective systems with an approximated limiting current of $I_{Tot}^{Lim} = 0.25 \pm 0.05$µA [Fig.3E]. When activating the microvalve with P=18psi (~60% closure), fluorescence clearly showed ICP-related depletion and preconcentration plug development at its edge [Fig.3B]. Concurrently, the chronoamperometric measurement showed current reduction over time in conjunction to visualization of the continuous growth of the depletion layer [Fig.3C-D]. A further increase in the pressure (>21psi) resulted in faster development of the depletion layer and a larger current reduction. In contrast to a lower P, for P>21psi, the microvalve deformed with sub-micron gaps and blocked >92% of the main channel. The gap size was estimated by the fact that 1µm polystyrene GFPs failed to pass through generating a steric-based filtration (Fig.S6). Such a considerable closure resulted in a significantly decreased flow rate (down to ~8µm/sec) and a negligible preconcentration factor (Mode [III]). Application of P in the range of 13-16psi resulted in an unstable ICP that appeared and disappeared in an uncontrolled manner. Taken together, a semi-closed microvalve with P≈18±1psi provided the best conditions for robust and an effective molecule preconcentration. Of note, the cross-section reconstruction and the particles trajectories suggested that starting from 18psi, the ceiling of the main microchannel underneath the microvalve had already partially collapsed onto the bottom surface at the center of the main channel. The partial contact between the ceiling and the bottom surfaces of the microchannel may have resulted in the formation of nano-channels[28–30]. However, the contribution of these nano-channels to the developed ICP was ruled out by repeating the same experiments in an identical microchannel system without a CEM coating. Even for the maximum tested P (32psi), there was no indication of ICP, likely due to lack of ionic permselectivity of the obtained nanochannels at this relatively strong electrolyte ionic strength (~10mM) [Fig.S7]. The classification of the system response into one of the three modes depends not only on the applied P but also on the applied voltage drop.

Increasing/decreasing the voltage drop shifted the transition between the under-limiting and over-limiting current regimes to a lower/higher P due to a higher/lower $I_{Tot}$ that reached above/below $I_{Tot}^{Lim}$, respectively [Fig.S8].

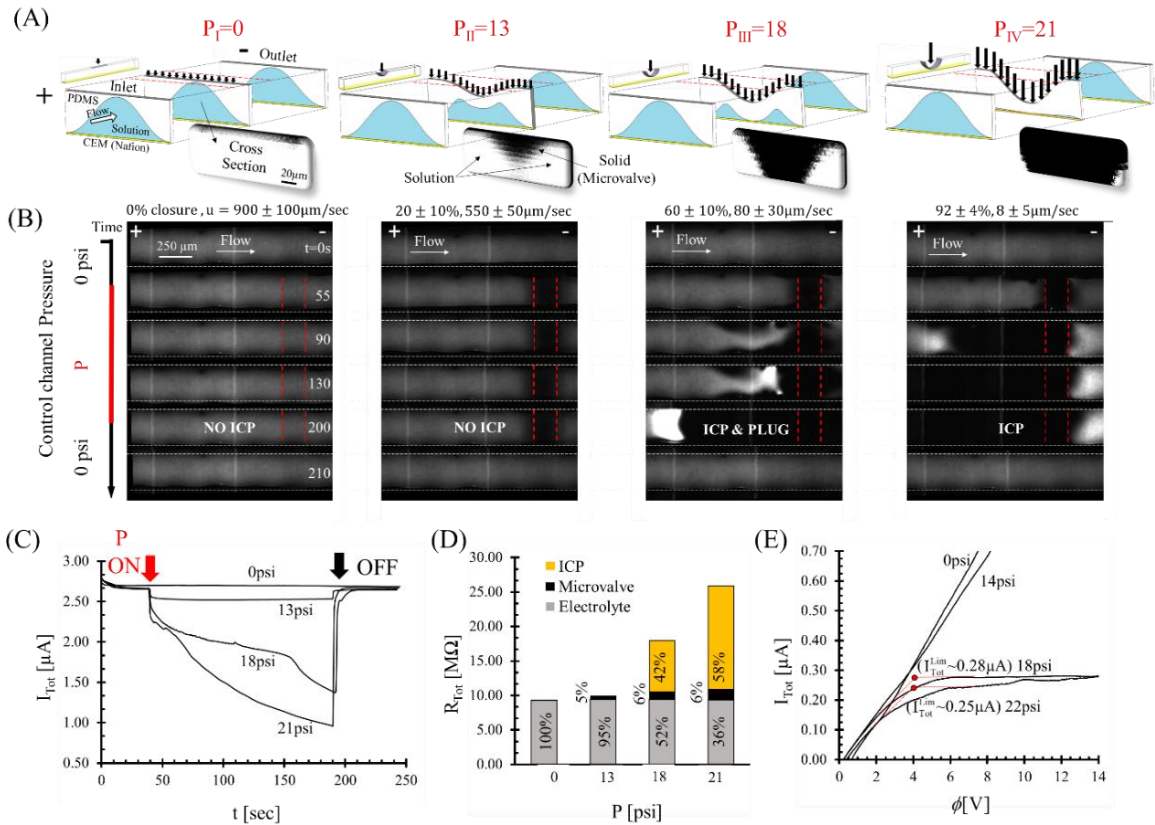

**Figure 3: Experimental results of the effect of microvalve deformation level on the electrically driven ion transport response.** (A) Schematic description of the main microchannel deformation, in addition to optical reconstruction of the microchannel cross-section underneath the center of the microvalve (black and white colors represent solid surfaces, i.e., microvalve wall, and the fluid, respectively). (B) Fluorescence intensity over time in a system with a deactivated (P=0psi at t<45s and t>190 s) versus activated microvalve (varied P at 45<t<190s). Each column represents a different control channel pressure (P) which gradually increases from left (0psi) to right (21psi). The percentage of microvalve closure (estimated from (A)) and the average flow velocity (u, calculated using GFPs, Fig.S6) are indicated. A constant voltage drop (25V) was applied throughout the entire operation period (0<t<250s). (C) Chronoamperometric measurements of the total current response ($I_{Tot}$) in correspondence to the examined conditions in (B). (D) Assessment of the relative contribution of each resistance component on the overall electrical resistance ($R_{Tot}$) measured in (C). The resistance components are the Ohmic electrolyte (gray), Ohmic microvalve deformation (black), and ICP-related depletion layer (yellow). (E) Current-voltage response (scan rate 7.5mV/sec) with the estimated $I_{Tot}^{Lim}$ (red dot).

*ICP dependency on microvalve deformation – numerical simulations investigation*

A study case of 94% local closure of the microchannel (from its initial open state, $\tilde{\phi}=75$, $\tilde{u}=65$) was numerically analyzed (Fig.4A-B). The electrolyte ion concentrations ($\tilde{c}_\pm$) and the electrical field ($\tilde{e}$) distributions indicated a local generation of ICP starting at the narrowed cross-section region ($\tilde{x}=1$) representing the activated microvalve. Plotting the analyte concentration ($\tilde{c}_A$) as well demonstrated the preconcentration plug development over time. At time zero ($\tilde{t}=0$), when voltage drop was applied ($\tilde{\phi}=75$), both the electrolyte ions and analyte were initially uniformly distributed within the microchannel. Then, although the CEM layer (marked as a yellow rectangle) was uniformly coated on the bottom surface, at $\tilde{t}>0$ depletion/enrichment layers of ions were generated only from the anodic/cathodic sides of the narrowed cross-section region (Fig.4A). The ionic current streamlines showed that the closure diverted the current streamlines from the solution towards the CEM. Within the depletion layer, the localized high electric fields induced strong electrophoretic forces on the negatively charged analyte molecules (Fig.4B), which, together with the counteracting background advection, resulted in accumulation of the analyte into a plug. Numerical calculation of ionic current ($\tilde{j}_{Tot}$) as a function of time exhibited a current reduction, in agreement with the experimental results [Fig.4C]. Furthermore, the ionic currents passing through each region separately underneath the microvalve ($\tilde{j}_{Sol}$ and $\tilde{j}_{CEM}$) were examined for each level of closure. With an open channel (0% closure), $\tilde{j}_{CEM}$ approached ~10% of $\tilde{j}_{Tot}$, and gradually increased to >70% for a channel with 96% closure (Fig.4D). In terms of ICP, the current-voltage curves showed that the limiting and the over-limiting regimes were reached at lower $\tilde{j}_{Tot}$ for higher degrees of closure (Fig.4E). Thus, tuning the closure adjusts the ratio of $\tilde{j}_{CEM}$ to $\tilde{j}_{Tot}$, which is responsible for triggering the over-limiting ICP at lower $\tilde{j}_{Tot}^{Lim}$. Accordingly, an analyte plug was only formed after crossing the limiting regime (Fig.4F). For example, application of a constant voltage of $\tilde{\phi}=78$ fell within the under-limiting (Ohmic-like) regime for 90% closure without formation of a plug (Mode [I]). Increasing the closure to 94%, while maintaining the same voltage, switched the response to the over-limiting regime and provided the conditions required for plug formation (Mode [II]). However, although further closure of the cross-section led to decrease of $\tilde{j}_{Tot}^{Lim}$, slower growth of the preconcentration factor was obtained due to the reduced flow rate.

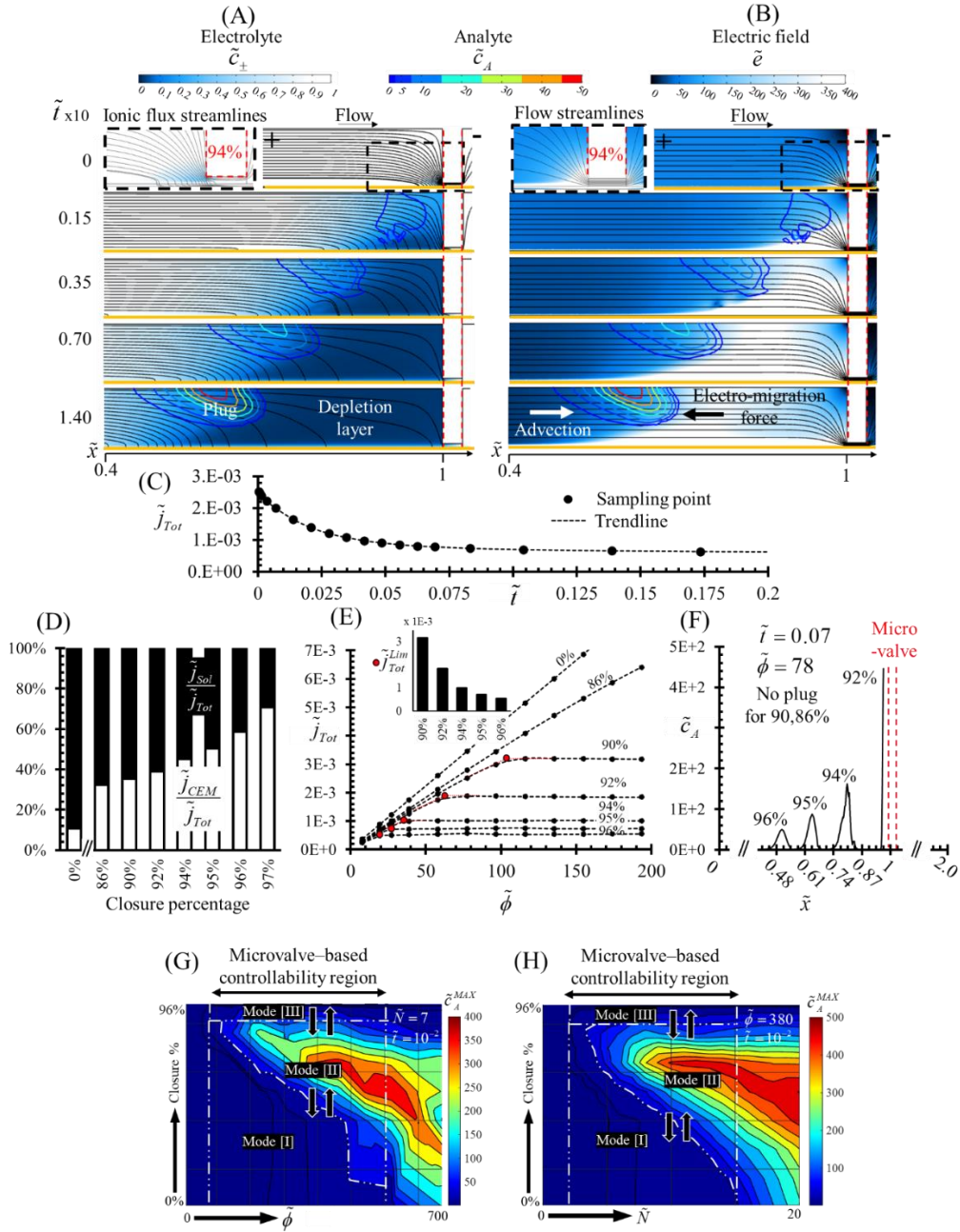

**Figure 4: Numerical simulations investigation of the effect of local closure percentage on the electrically driven ion transport response.** A study case of 94% main channel closure (applied $\tilde{\phi} = 78$, $\tilde{u} = 65$) is presented, wherein the membrane is depicted as a yellow rectangle and has $\tilde{N} = 7$. (A) Electrolyte concentration ($\tilde{c}_\pm$) distribution over non-dimensional time ($\tilde{t} = 0$, 0.015, 0.035, 0.07, 0.14), wherein the black lines indicate the electrical current streamlines, and the colored contour lines denote the analyte concentration ($\tilde{c}_A$). (B) Electric field ($\tilde{e}$) distribution, flow streamlines (black lines) and analyte concentration (colored contour) for the same conditions as (A). (C) Ionic current ($\tilde{j}_{Tot}$) response over time.

The effect of closure percentages on: (D) the initial ($\tilde{t}=0$) ratio of $\tilde{j}_{CEM}$ and $\tilde{j}_{Sol}$ to that of $\tilde{j}_{Tot}$ (black lines). (E) Ionic current-voltage response. For each curve, $\tilde{j}_{Tot}^{Lim}$ is marked with a red dot. (F) Analyte plug location and preconcentration factor under $\tilde{\phi}=78$ at $\tilde{t}=0.07$. The maximum analyte concentration ($\tilde{c}_A^{MAX}$) obtained at given time of $\tilde{t}=10^{-2}$ within the main channel as a function of (G) $\tilde{\phi}$ (examined range of $0<\tilde{\phi}<700$, $\tilde{N}=7$, $\tilde{u}=65$) or (H) $\tilde{N}$ ($\tilde{\phi}=380$, $0<\tilde{N}<20$, $\tilde{u}=65$) for various closure percentages (0-96%). The response phase diagram is divided into regions (dashed white lines represent the boarders) that enable microvalve-based controllability over the plug generation: Mode[I]-Ohmic, [II]- ICP with a plug, [III]-ICP with a negligible plug.

*Optimized conditions for effective ICP-driven preconcentration.*

The above described local microvalve-based controllability of the ICP and its associated preconcentration plug was not always achieved as it necessitates certain conditions as detailed herein. Previous works [e.g., Park et al.[19] and Kim et al.[20]] using a fixed microchannel configuration resulted in ICP and plug generation only after a sufficiently large electric potential difference was applied. Thereby, to enable dynamic controllability over the ICP and its plug without changing the constantly applied electric potential difference, $I_{Tot}$ must kept below $I_{Tot}^{Lim}$ for an opened microchannel. This can be achieved by lowering the initial applied voltage drop. Alternatively, $I_{Tot}^{Lim}$ can be increased by lowering the ion permselectivity of the CEM (Fig.S9). Increasing the electrolyte ionic strength resulted in a decrease in $N$ and with it, a decrease in the ion permselectivity[33]. While application of a constant electric current ($I_{Tot}$ =2.5µA) did not generate ICP, for a relatively strong ionic strength (10mM) without activating a microvalve, a plug was generated with a weaker ionic strength (1mM) under the same applied current. The resulted in a plug positioned at the external edge of the CEM coating closest to the anode reservoir despite the fact that the microvalve was deactivated. Numerical simulations of the system response (Fig.4G-H), depicted in terms of maximum achievable analyte concentration ($\tilde{c}_A^{MAX}$), for various $\tilde{N}$ (0-20), $\tilde{\phi}$ (0-700), and closure percentages (0-95%), determined the $\tilde{N}$ or applied $\tilde{\phi}$ below which a plug cannot be generated regardless of the closure percentage, due to the lack of induced ICP ($\tilde{j}_{Tot} < \tilde{j}_{Tot}^{Lim}$ for all closures). On the other hand, increasing $\tilde{N}$ or $\tilde{\phi}$ beyond a certain threshold resulted in the permanent generation of a plug even though the microchannel is fully open ($\tilde{j}_{Tot} > \tilde{j}_{Tot}^{Lim}$ for all closures). Between these two limits, the closure percentage can be modified to control plug generation by switching the response mode (Mode [I],[II], and [III]). For each $\tilde{N}$ or $\tilde{\phi}$, the closure percentage that provides for the highest preconcentration factor obtained within a given time interval ($\tilde{t} = 10^{-2}$) can be found. It can be seen that the closure percentage needed to reach this optimum point of highest preconcentration factor is reduced with increasing $\tilde{N}$ or $\tilde{\phi}$.

*Plug manipulation using an array of microvalves*

After demonstrating the potential to dynamically control ICP and its associated plug using a single microvalve, controllability was extended to an array of microvalves. Owing to the uniform CEM coating, the number of plugs and their locations can be dynamically determined. While the choice of which microvalves are activated determines where ICPs are generated, the interaction between two adjacent ICPs prevents further propagation of the plug and sets its final location. The microvalves within the array can be operated simultaneously or in a certain sequence, dictated by the desired application. Such microvalve-based programmability replaces the need to pattern the CEM coating, e.g. into an array of individually addressable CEM pairs[19]. For example, dynamic operation sequence of two microvalves enables control over the number of plugs formed and later merging them to a single plug, as shown in Fig.5A. The number of plugs can be increased by activating more microvalves (Fig.S10). Yet, each activated microvalve contributes to the overall hydrodynamic resistance, and, therefore activation, of too many microvalves leads to a low accumulation rate of molecules and a significantly lower preconcentration factor of each plug. In addition to the use of microvalves for controlled ICP generation, we demonstrated the ability to capture an analyte plug between two activated microvalves long after turning the electric field off and with it, loss of the ICP (Fig.5B). Without such physical isolation of the plug, as soon as the electric field is turned off, the balance between the electro-migration and advection is disturbed with the plug being advected downstream (Supp. M3). The ability to form a stagnant plug without the need for a continuously applied electric field is of importance for various applications, e.g., electrochemical and immunoassay sensing as well as analyte separation.

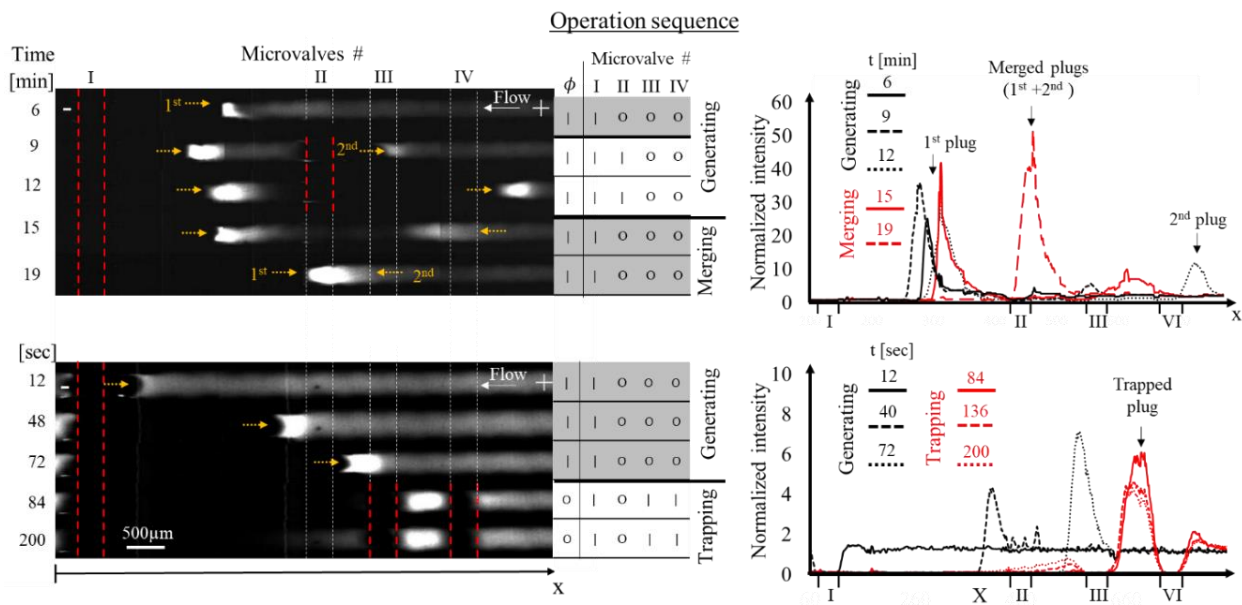

**Figure 5: Experimental dynamic operation of multiple microvalves (i.e., multiple control channels pressurized independently with $P_I$, $P_{II}$, $P_{II}$ and $P_{IV}$) connected in series.** Using the operation sequence shown in the tables, the external voltage drop was turned on (marked as '|', $\phi = 20V$) and off (marked as 'O', $\phi = 0V$), and the microvalves were separately activated ('|', P=19psi) and deactivated ('O',P=0psi). (A) Programmable operation of generation (0<t<75s) and merging (t>75s) steps of two preconcentrated analyte plugs using two microvalves, by activating $P_I$ and $P_{II}$. (B) Generation of a single plug (0<t<22s, by activating $P_I$) and its trapping between two upstream microvalves (t>22s, by activating $P_{III}$ and $P_{IV}$), while turning off $\phi$. The fluorescence intensity profiles at select time points for each operation are presented on the right.

## CONCLUSION

To summarize, we have developed a method to tune the effective ion-perm selectivity and hydraulic permeability of a microfluidic system for on-demand ICP and preconcentration of a target analyte. By integrating a single microvalve on top of a microchannel with uniform CEM coating at its bottom, we have achieved a wide range of responses starting from a non-selective to highly selective ion transport behaviors. We have shown that controlling the deformation of the microvalve also enables to dynamically control the interplay between the CEM and the solution resistances in the deformed microvalve region, thereby, to control the ionic current that passes through the CEM. However, increasing the microvalve deformation results in increased hydrodynamic resistance and decreased flow rate and accumulation of analytes within the plug. Hence, owing to the unique tuning capabilities mentioned above, we have found the optimal deformation level which enabled sufficient ion transport through the CEM for generation of ICP, while keeping a sufficiently large solution flow throughput for an effective and high concentration factor of the associated preconcentration analyte plug. The deformation level can be retuned for any change in the operation condition and thus enabling an efficient operation without the need of redesigning and fabricating a new chip. Additionally, application of multiple microvalves in series introduces a robust and simple way to implement a variety of new functionalities into lab-on-a-chip devices, e.g., programable manipulation of multiple preconcentration plugs for sensitive multiplex sensing, suppression of biofouling as well as plug dispersion, while maintaining the well-known application of microvalves as steric filtration.

**METHODS**

*Experimental setup*

The chip consisted of a main microchannel (35-mm-long, 250-µm-wide and 55-µm-high) uniformly coated with a CEM (Nafion$^{TM}$) at its bottom surface (CEM thickness of ~2% of the microchannel height), and with several microvalves (3-7) embedded along the length of the top of the main microchannel. Each microvalve was individually controlled by a control channel (orthogonal to the main channel, 230-µm-wide, 320-µm-high) filled with a pressurized fluid (deionized water, DI). Pressurization of the fluid inside the control channel deforms the ceiling of the main channel, which reduces its cross-sectional dimension. Precise control of the main channel cross-section can thereby be achieved by tuning the pressure inside the control channel (P). All channels were comprised of 1:10 (base:cross-linker) Polydimethylsiloxane (PDMS,) , while a deformable thin PDMS film (~60-µm-high) separating the bottom part of the control channel from the top part of the main channel (i.e., no flow crosses between the two channels). The fabrication process, chip geometry and the experimental setup, including the pressure control system, are fully described in Supplementary Figures S1-4. Two silver-silver chloride electrodes (Ag/AgCl, A-M system, 0.015" diameter) were immersed within the cathodic and anodic reservoirs at the opposite ends of the main microchannel to apply either a voltage drop or an electrical current using a source-meter (Keithley 2636). A low concentration analyte (relative to the electrolyte ion concentrations) of negatively charged fluorescent dye (Alexa 488, Thermo Scientific Inc.) was used as the target bioparticle for ICP-driven preconcentration within a 10mM KCl aqueous electrolyte (1.6 mS/cm, pH 7, unless otherwise is mentioned). The fluorescence intensity of the dye, visualized and captured using a spinning disk confocal system (Yokogawa CSU-X1), inverted microscope (Eclipse Ti-U, Nikon) and a camera (Andor iXon3), was analyzed by normalizing the local fluorescent dye intensity by that of the initial intensity measured before electric field application and microvalve activation. The net flow was driven by a pressure difference between the reservoirs, and the velocity was measured by monitoring 1µm polystyrene (green fluorescent protein (GFP) labeled particles, Thermo Scientific Inc.).

*Numerical simulations*

The fully coupled Poisson-Nernst-Plank (PNP) equations were solved along with the simplified Navier-Stokes equation (neglecting inertia and body forces) for an incompressible fluid using a two-dimensional (2D) time-dependent model (COMSOL Multiphysics 5.3). A microchannel (2L-long) with a CEM embedded at the bottom surface (thickness of 2% of the microchannel height), initially filled with a symmetric binary ($z_\pm = \pm 1$) electrolyte of equal ions diffusivities ($D = D_\pm$) and low analyte concentration ($c_{A,0}$, relative to the concentration of the dominating background electrolyte ions, $c_{\pm,0}$), was simulated. The following normalization was used to present the results: axial coordinate $x = L\tilde{x}$, ion concentration $c_i = c_{i,0}\tilde{c}_i$, electrical potential $\phi = (RT/F)\tilde{\phi}$, electric field $E = (RT/FL)\tilde{e}$, ionic flux $j_i = (Dc_{i,0}/L)\tilde{j}_i$, average flow velocity $u = (D/L)\tilde{u}$, and time $t = (L^2/D)\tilde{t}$. Here, tilde notations denotes nondimensional parameters, subscript $i$ denotes the different electrolyte ($i = +, -$, for positively and negatively charged ions, respectively) and analyte ($i = A$) ion species within the solution, $R$ is the universal gas constant, $T$ temperature, and $F$ the Faraday number. A fixed volumetric charge density ($N = c_0\tilde{N}$) was defined within the CEM[31,32], along with a solvent impermeability condition at the CEM-microchannel boundaries. To account for microvalve actuation, the microchannel cross-section was narrowed (0.05$L$ long) at its center. For simplification, only a single microvalve was considered, and the shape of the narrowed section was modeled as a rectangular gap of uniform height that approximates the average realistic non-uniformly deformed gap height. The different deformation levels were implemented by tuning the closure percentage of the microchannel. From an open microchannel with 0% closure for the undeformed microvalve, up to an almost completely closed microchannel with 98% closure for the fully deformed microvalve. For additional information, see Supplementary Fig.S5.

## ASSOCIATED CONTENT

### Supporting Information

The Supporting Information is available free of charge at …

    Figures S1-S10, description of the numerical simulations, and movies

## AUTHOR INFORMATION

### Corresponding Author

    **Gilad Yossifon** – School of Mechanical Engineering, Tel-Aviv University, Israel; Email: gyossifon@tauex.tau.ac.il

### Author

    **Barak Sabbagh** - Faculty of Mechanical Engineering, Technion–Israel Institute of Technology, Israel

    **Sinwook Park** - School of Mechanical Engineering, Tel-Aviv University, Israel

## ACKNOWLEDGEMENTS

We acknowledge the support of the Israel Science Foundation (ISF 1934/20). We wish to acknowledge the Technion Russel-Berrie Nanotechnology Institute (RBNI) and the Technion Micro-Nano Fabrication Unit (MNFU) for their technical support.

*Nano Lett.* **2020**, *20*, 8524–8533. https://doi.org/10.1021/acs.nanolett.0c02973.